\documentclass[final,5p,times,twocolumn]{elsarticle}

 \usepackage{graphics}
\usepackage{amssymb}
\journal{Computer Physics Communications}

\begin{document}

\begin{frontmatter}



\title{Crowd dynamics - being stuck}


\author{ Przemys{\l}aw Gawro\'nski and Krzysztof Ku{\l}akowski}

\address{Faculty of Physics and Applied Computer Science, AGH University of Science and Technology, al. Mickiewicza 30, PL-30059 Krak\'ow, Poland}

\begin{abstract}
We consider a crowd of $N$ persons trying to exit some area trough a small exit.  The probability is calculated that an individual is able to withdraw from the crowd under one's own steam. The problem is simulated within the generalized force model (D. Helbing et al., Nature 407 (2000) 487), and all model parameters are taken from this paper. The results indicate, that in a crowd of 150 persons, this probability is not greater than ten percent. We also evaluate the number of helpers necessary to get the above probability of fifty percent.
\end{abstract}


\end{frontmatter}

% \linenumbers


\section{Introduction}

When applications of natural sciences to human beings are considered, the problem of unpredictability of human mind is an eternal motif.
In psychology, neurophysiological arguments overlap with philosophical ones \cite{damasio,gluck}. In sociology, the empiricistic
foundations of natural sciences have been undermined by interpretative sociology of Weber and Simmel \cite{oxford}; 'Determinism is dead' is a catchphrase of today \cite{goeteborg}. Against this background, the modeling of crowd dynamics - which is our aim here - could seem to be an extreme example of a mechanicistic reductionism. On the other hand, prediction is an ultimate aim of all sciences; here the famous statement on scientific aim 'Savoir pour pr\'evoir et pr\'evoir pour pouvoir' by Auguste Comte remains valid and desirable \cite{alridge}. In fact, a crowd can be seen as a many-body system with local interactions; in such systems, statistical laws should allow for some predictions. Once a comparison with experimental data became possible \cite{dh3,dh4}, the critique from the hermeneutically oriented audience is less convincing.\\

Modeling of the crowd dynamics is known for more than 50 years \cite{1958}. The methods have been advanced much in 90's by Dirk Helbing and cooperators; reviews and lists of references can be found in \cite{dh1,dh2,xia}. Among the methods, the social force model seems to be most realistic \cite{nature}. This is a set of differential equations of motion, where positions and velocities of pedestrians are time-dependent variables; the approach is equivalent to the molecular dynamics, where human desires are encoded in the form of social forces. Within this model, a number of problems have been addressed, as lane formation, strip formation, turbulent waves, herding and bottlenecks \cite{dh2}. In this paper we address more directly to the problem of the phase of a simulated crowd. Namely, we ask for the conditions when an individual or a small group can change their position with respect to their neighbours in the crowd. If they are stuck, we refer to a clogged phase. This criterion is a direct analogy to the Monte Carlo simulations of the crystallization phase transition of hard spheres \cite{kranendonk,biben}. We note that although the dynamics of hard spheres is much simpler than the crowd dynamics, systematic numerical studies of the finite size effect have been possible only recently \cite{binder}. \\

The scenario to be simulated here within the social force model is as follows. Pedestrians numbered by $i=1,...,N$ are going to leave a room through a small exit. During this process, we monitor the sum $S_i$ of mechanical compressive forces acting on each individual. Once
this sum exceeds some prescribed value $S_c$ for any individual $j$, the direction of the desired motion of this individual is reverted from the vector towards the exit to the opposite. Further, a number $K$ of pedestrians who are nearest neighbours of $j$-th one decide to 
accompany her/him. Then their directions of desired motion are set equal to the direction of $j$-th individual, and their social forces towards $j$-th individual, initially repulsive, change signs. The outcome of the simulation is the probability $P$, that the crowd throws $j$ out through the exit, despite her/his struggling to withdraw from the crowd. As $K$ pedestrians help $j$, $P$ is expected to decrease with $K$.\\

Among the problems considered by other authors, this scenario is somewhat similar to the bi-directional flows in the bottleneck problem \cite{dh2}, where two flows of pedestrians walking in opposite directions met at a narrowing of the path. Also, the effect of clogging has been demonstrated in \cite{buz,par1} as long time periods when nobody crosses an exit. In \cite{hemo}, a picture is drawn (Figs. 4 and 5) where a narrow queue of pedestrians is formed through the crowd of individuals. The effect of the crowd solidification as opposed to the lane formation was discussed in \cite{hfv}. In this paper, the phase of coherent motion in lanes was destroyed by an added noise. In \cite{par2}, the authors defined clusters of individuals as sets of those who interacted via physical forces. Above some optimal value of the desired velocity, the distribution of size of these clusters was found to be strongly modified.  In our earlier paper \cite{gawr} we discussed a similar scenario, where pedestrians around make more place to a handicapped person, for she/he can leave the room.  In the text presented here, a single individual or a small group intends to move with respect to the crowd; then, the crowd itself is considered as a medium for individual pedestrians or their small groups. Up to our knowledge, this issue has not been discussed yet. \\

The paper is organized as follows. In the next section the social force model is explained in more detail. We adopt the formulation and the values of the parameters used in \cite{nature}. Section 3 provides the numerical results on the probability $P$ of being stuck as dependent on the model parameters $N$, $K$ and $S_c$. Last section is devoted to concluding remarks.

\section{The model}

The model equations of motion are adopted from the generalized force model \cite{hemo,hfv,nature}. Such an equation for a person of mass $m$ is as follows

\begin{equation}
m\frac{d\mathbf{v}_i}{dt}=m\frac{\mathbf{v}(\mathbf{r}_i)-\mathbf{v}_i}{\tau}+\sum_{j(\ne i)}\mathbf{f}_{ij}+\sum_W\mathbf{f}_{iW}
\end{equation}
where the first term on right hand side is the tentative acceleration of a person $i$ who intends to have the velocity $\mathbf{v}(\mathbf{r}_i)$,
dependent on the coordinates $\mathbf{r}_i$; as a rule, the vector $\mathbf{v}$ points to the exit center (large distance from the person to the exit) 
or to the closest point of the exit (small distance). In our simulations, the absolute value of the desired velocity $\mathbf{v}(\mathbf{r}_i)$ is 3 m/s; this is purposefully higher than the optimal value 1. 375 m/s \cite{par2}. Further, $\tau$ is the characteristic time of this acceleration, $\mathbf{v}_i$ is the actual velocity of $i$-th person, $\mathbf{f}_{ij}$ is the force exerted on $i$-th person by $j$-th person, and $\mathbf{f}_{iW}$ is the force exerted on $i$-th person by a wall $W$. The force $\mathbf{f}_{ij}$ contains three components; 'psychological' interaction which describes the tendency of $i$ and $j$ to keep distance between each other, and two physical interactions between their bodies: radial force and slide friction. The psychological interaction is equal to $A_i\exp((2R-\big\|\mathbf{r}_i-\mathbf{r}_j\big\|)/B$), where $r_i$ is the position of $i$-th person, $R$ is the mean 'radius' of the vertical projection of the human body. This psychological part of $\mathbf{f}_{ij}$ will be modified in our simulation. The radial physical force is equal to $kg(2R-\big\|\mathbf{r}_i-\mathbf{r}_j\big\|)$, where $g(x)=x$ if $x>0$, $g(x)=0$ elsewhere, and $k$ is a constant. The physical friction is assumed to be
$\kappa g(2R-\big\|\mathbf{r}_i-\mathbf{r}_j\big\|)((\mathbf{v}_j-\mathbf{v}_i)\cdot \mathbf{t_{ij}})\mathbf{t_{ij}}$, where $\mathbf{t_{ij}}$ is the unit vector of tangential direction to the body surfaces. The same expressions of the physical forces are used to describe the body-wall interaction. The instant values of the velocities $\mathbf{v}_i$ allow to update the positions $\mathbf{r}_i$ as well. The parameters of the system are adopted from \cite{nature}. Namely, $A$ = 2000 N, $B$ = 0.08 m, $\tau$ = 0.5 s, $k$ =1.2 $\times 10^5$ $kg/s^2$, $\kappa/k = 2$, $R$ = 0.3 m and $m$ = 75 kg.  The room is $30\times20 m$, with the exit of width of $1 m$ in the middle of the shorter wall. The initial positions of the agents are selected randomly, but the initial overlaps of them are excluded.\\

 \begin{figure}[ht]
 %\vspace{0.3cm}
 \centering
 {\centering \resizebox*{9cm}{6cm}{\rotatebox{-90}{\includegraphics{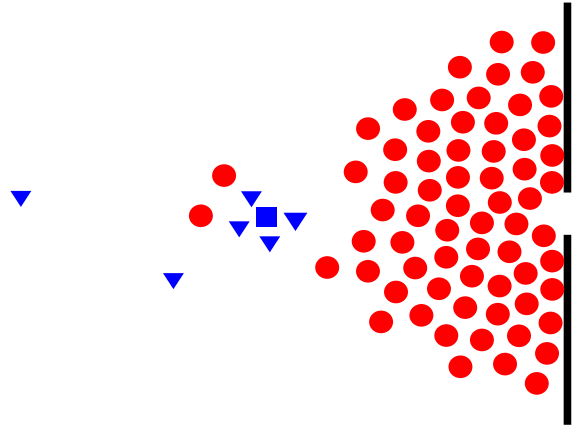}}}}
 %\vspace{0.3cm}
\caption{A spatial configuration of individuals near the exit. X is marked with square (blue online), the six helpers are marked with triangles (blue online), the others are with circles (red online). In this example, the help is successful. }
 \label{fig-1}
 \end{figure}

 \begin{figure}[ht]
 %\vspace{0.3cm}
 \centering
 {\centering \resizebox*{9cm}{6cm}{\rotatebox{-90}{\includegraphics{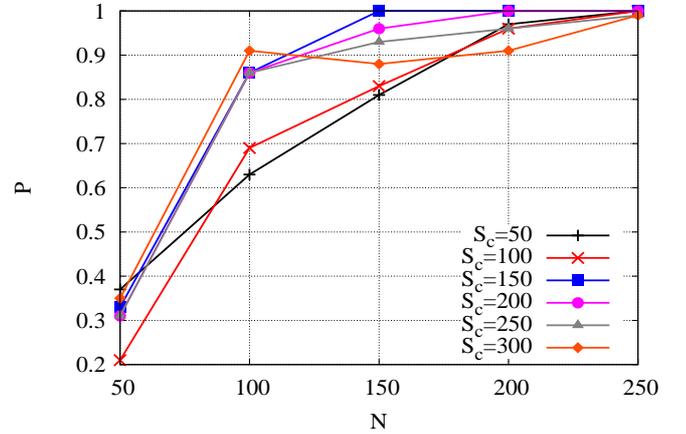}}}}
 %\vspace{0.3cm}
\caption{The probability $P$ of being thrown through the exit as dependent on the number of persons $N$ in the crowd, for different values of the threshold $S_c$. These data are obtained for $K$ = 0 (no helpers).}
 \label{fig-2}
 \end{figure}

To determine if a person is stuck or not, a numerical experiment is performed as follows. The sum $S_i$ of compressive mechanical forces acting on each individual $i$ is registered during the motion. Once for some individual X the sum $S_x$ exceeds some threshold value $S_c$ prescribed at the beginning of the simulation, the desired  direction of motion for this individual is reverted. Now this direction is not to the exit but the opposite one. We assume that all individuals act with the same force. We are also interested if a collective action of neighboring individuals could change the outcome of the experiment. To check this, we change also the desired direction of $K$ individuals, who are closest to X when the threshold value is exceeded. Their desired direction of motion is now equal to the desired direction of X. Also, the repulsive psychological forces between X and his neighbours change sign to be attractive. Now the group of $K+1$ individuals tries to evade the exit, as if they tried to help a victim of the interpersonal forces in the crowd. In Fig. 1, an example is shown for $K=6$, where the help is successful. In both experiments, if X crosses the exit despite this change of her/his intention, we call the crowd 'jammed'. \\

\section{Results}

The outcome is the probability $P$ that X is thrown out through the exit against her/his will. In Fig. 2 this probability is shown for $K$ = 0 (no helpers) against the crowd size $N$, for various values of the threshold value $S_c$. Each point on these results is an average over 70 samples. As we see, the results only weakly depend on $S_c$. On the contrary, the crowd size $N$ is decisive. As we see, the probability $P$
increases with $N$ from less than 0.4 for $N$ = 50 to about 1.0 for $N$ = 250.\\

The simulations are repeated in the presence of $K$ helpers, for $K$ between 1 and 10, $S_c$ = 150 N. The results are shown in Fig. 3. Each point is an average over 200 samples, and the error bars are the differences between the averages over first and second hundred of samples. Here again, $N$ is relevant, but the crowd size can be to some extent neutralized by the number of helpers. For example, $P$ close to 0.5 can be achieved in a crowd of $N$ = 100 persons with $K$ about 2 helpers, in a crowd of $N$ = 200 persons with $K$ about 5 helpers and so on. In simulations, we observed many times that the group of helpers become dispersed, as in Fig. 1.  In the presented results this dispersion is neglected; this means that the effective number of helpers is perhaps smaller.\\

 \begin{figure}[ht]
 %\vspace{0.3cm}
 \centering
 {\centering \resizebox*{9cm}{6cm}{\rotatebox{-90}{\includegraphics{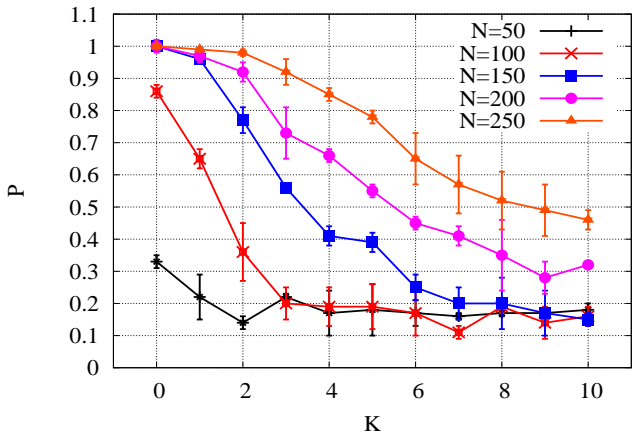}}}}
 %\vspace{0.3cm}
\caption{The probability $P$ of being thrown through the exit against the number of helpers $K$, for different values of the crowd size $N$. Here, the threshold force $S_c$ = 150 N. }
 \label{fig-3}
 \end{figure}

 \begin{figure}[ht]
 %\vspace{0.3cm}
 \centering
 {\centering \resizebox*{9cm}{6cm}{\rotatebox{-90}{\includegraphics{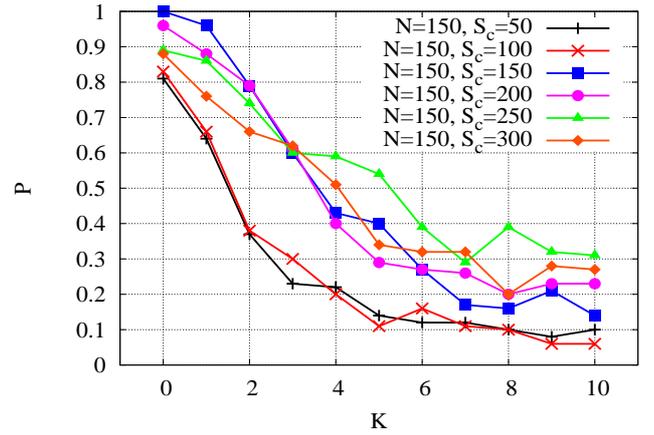}}}}
 %\vspace{0.3cm}
\caption{The probability $P$ of being thrown through the exit against the number of helpers $K$, for different values of the threshold force $S_c$. Here, the crowd size is $N$ = 150.}
 \label{fig-4}
 \end{figure}

The simulations with helpers are repeated also for $N$ = 150 and different $S_c$. These results are shown in Fig. 4. Each point is an average over 100 samples. These results confirm, that the threshold value $S_c$ does not influence much the probability $P$. We have performed also a similar experiment with the variation of the time parameter $\tau$. Other parameters of this experiment were $N$ = 150, $K$ = 0 and $S_c$ = 150 N. The obtained probability $P$, which is 1.0 for $\tau$ = 0.5, is not less than 0.95 for $\tau$ as large as 5.0. We deduce that the variation of $\tau$ is not relevant. \\

\section{Discussion}

Our numerical results indicate that once the crowd size $N$ exceeds 150-200 persons, it is unlikely that a single individual can withdraw under one's own steam. Then, any large gathering of people should be treated as a potentially dangerous medium. This conclusion is supported by the historical data on crowd disasters \cite{cddis}.  \\

In the clogged phase, the only chance to leave the crowd is to mobilize a group of helpers nearby. We note that once this group is dispersed, each separate person is helpless in the same way. What does matter in these conditions is the communication between people. The question arises, how many helpers must be found to have a chance of 50 percent to withdraw from the crowd. Let us denote this number as $K_{50}(N)$. Because of the complexity of the problem, the accuracy of our results allows to classify them as semi-quantitative only. We made an attempt to fit $log(K_{50})$ against $log(N)$ to obtain the exponent $\beta$ in the tentative scaling relation $K_{50}\propto N^\beta$. The result is that $\beta$ =1.88 $\pm$ 0.05. We feel entitled to claim that $\beta$ is larger than 1.0. However, as $K_{50}$ cannot be greater than $N$, this behaviour must end with some crossover for larger $N$. A more quantitative evaluation of the function shape of $K_{50}(N)$ needs much more computational power.\\
\begin{figure}[ht]
 %\vspace{0.3cm}
 \centering
 {\centering \resizebox*{9cm}{6cm}{\rotatebox{-90}{\includegraphics{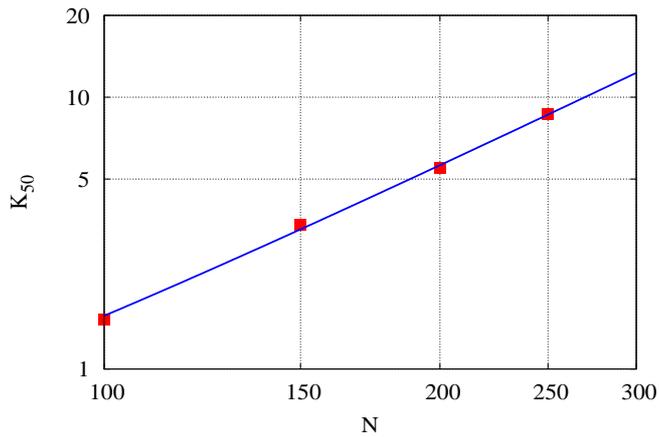}}}}
 %\vspace{0.3cm}
\caption{The probability $P$ of being thrown through the exit against the number of helpers $K$, for different values of the threshold force $S_c$. Here, the crowd size is $N$ = 150.}
 \label{fig-5}
 \end{figure}

Summarizing, in a crowd of some hundreds of people the difference between an autonomous human being and a piece of passive body is less than we would like to admit. Although at the exit the individuals leave the room one by one, in the middle of the crowd their mutual positions cannot be changed. In these conditions, the unpredictability of the human mind does not influence the trajectory of its owner. The obtained results should be helpful to evaluate human resources which are needed to tackle emergency situations in large gatherings of people.\\

\section*{Acknowledgements} The research is partially supported within the FP7 project SOCIONICAL, No. 231288.

\end{document}